\documentclass[twocolumn,preprintnumbers,amsmath,amssymb,superscriptaddress]{revtex4}

\usepackage{graphicx}
\usepackage{dcolumn}
\usepackage{bm}
\usepackage{times,mathptmx}
\usepackage{color}
\usepackage{hyperref}
\usepackage{placeins}

\usepackage{remreset}

\begin{document}

\title{A microchip optomechanical accelerometer}

\author{Alexander G.\ Krause}
\thanks{These authors contributed equally to this work}
\author{Martin Winger}
\thanks{These authors contributed equally to this work}
\author{Tim D.\ Blasius}
\thanks{These authors contributed equally to this work}
\affiliation{Thomas J.\ Watson, Sr., Laboratory of Applied Physics, California Institute of Technology, Pasadena, CA 91125, USA}
\author{Qiang Lin}
\affiliation{School of Engineering and Applied Sciences, University of Rochester, Rochester, New York 14627, USA}
\author{Oskar Painter}
\email{opainter@caltech.edu}
\homepage{http://copilot.caltech.edu}
\affiliation{Thomas J.\ Watson, Sr., Laboratory of Applied Physics, California Institute of Technology, Pasadena, CA 91125, USA}

\date{\today}

\begin{abstract}
The monitoring of accelerations is essential for a variety of applications ranging from inertial navigation to consumer electronics~\cite{krishnan_reviews_2007}. The basic operation principle of an accelerometer is to measure the displacement of a flexibly mounted test mass; sensitive displacement measurement can be realized using capacitive~\cite{acar_experimental_2003,kulah_noise_2006}, piezo-electric~\cite{tadigadapa_piezoelectric_2009}, tunnel-current~\cite{cheng-hsien_liu_characterization_1998}, or optical~\cite{krishnamoorthy_-plane_2008,zandi_-plane_2010,noell_applications_2002,berkoff_experimental_1996} methods. While optical readout provides superior displacement resolution and resilience to electromagnetic interference, current optical accelerometers either do not allow for chip-scale integration~\cite{krishnamoorthy_-plane_2008} or require bulky test masses~\cite{zandi_-plane_2010,noell_applications_2002}. Here we demonstrate an optomechanical accelerometer that employs ultra-sensitive all-optical displacement read-out using a planar photonic crystal cavity~\cite{eichenfield_picogram-_2009} monolithically integrated with a nano-tethered test mass of high mechanical $Q$-factor~\cite{verbridge_high_2006}. This device architecture allows for full on-chip integration and achieves a broadband acceleration resolution of $10\ \mathrm{\mu g/\sqrt{Hz}}$, a bandwidth greater than 20~kHz, and a dynamic range of 50~dB with sub-milliwatt optical power requirements. Moreover, the nano-gram test masses used here allow for optomechanical back-action~\cite{kippenberg_cavity_2007} in the form of cooling~\cite{genes_ground-state_2008} or the optical spring effect~\cite{Corbitt2007,lin_mechanical_2009}, setting the stage for a new class of motional sensors.
\end{abstract}

\maketitle

\begin{figure*}[t]
\includegraphics[width=2\columnwidth]{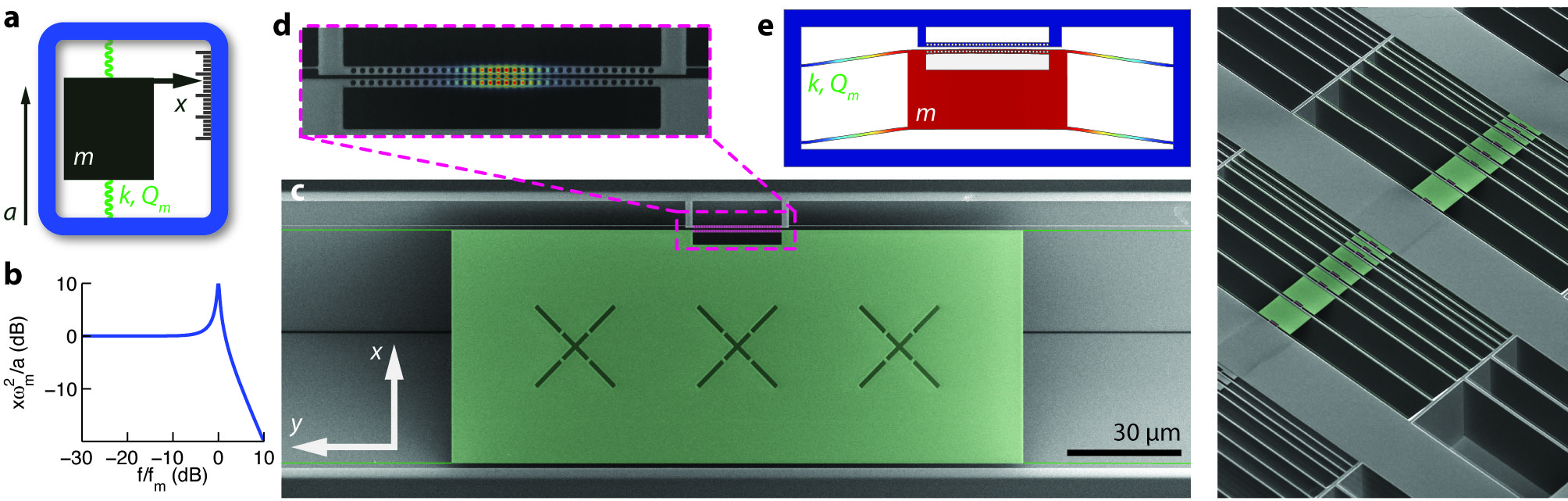}
\caption{\label{Fig1} \textbf{Overview of the accelerometer design.}
	\textbf{a}, Canonical example of an accelerometer. When the device (blue frame) experiences a constant acceleration $a$, a test mass $m$ undergoes a displacement of $x = ma/k$.
	\textbf{b}, Frequency response $|\chi(\omega)|$ of an accelerometer on a log-log plot featuring a resonance at $f_m = \sqrt{k/m}/2\pi$ with $Q_m = 10$.
	\textbf{c}, False-colored SEM-image of a typical optomechanical accelerometer. A test mass of size $150\ \mathrm{\mu m} \times 60\ \mathrm{\mu m} \times 400\ \mathrm{nm}$ (green) is suspended on highly stressed 150~nm wide and $560\ \mathrm{\mu m}$ long SiN nano-tethers, which allow for high oscillator frequencies ($>27\ \mathrm{kHz}$) and high mechanical $Q$-factors ($>10^6$). On the upper edge of the test mass, we implement a zipper photonic crystal nanocavity (pink). The cross-shaped cuts on the test mass facilitate undercutting the device.
	\textbf{d}, Zoom-in of the optical cavity region showing the magnitude of the electric field $\left| \mathbf{E}(\mathbf{r}) \right|$ for the fundamental bonded mode of the zipper cavity. The top beam is mechanically anchored to the bulk SiN and the bottom beam is attached to the test mass.
	\textbf{e}, Schematic displacement profile (not to scale) of the fundamental in-plane mechanical mode used for acceleration sensing.
	\textbf{f}, SEM-image of an array of devices with different test mass sizes.}
\end{figure*}

Due to the rapid development of silicon micro machining technology, MEMS accelerometers have become exceedingly popular over the last two decades~\cite{krishnan_reviews_2007}. Evolving from airbag deployment sensors in automobiles to tilt-sensors in cameras and consumer electronics products, they can now be found in a large variety of technological applications with very diverse requirements of their performance metrics. While sensors for inertial navigation systems require low noise levels and superior bias stability~\cite{zwahlen_navigation_2010}, large bandwidth is crucial for sensors in acoustics and vibrometry applications. However, there is a fundamental tradeoff between noise performance and bandwidth which can be understood from the basic operation principle of an accelerometer, illustrated in Fig.~\ref{Fig1}a.  When subjected to an acceleration $a(\omega)$ at frequency $\omega$, a mechanically compliant test mass experiences a displacement $x(\omega) = \chi(\omega) a(\omega)$ proportional to the mechanical susceptibility $\chi^{-1}(\omega) = \omega_m^2 - \omega^2 + i\frac{\omega\omega_m}{Q_m}$. Here, $\omega_m = 2\pi f_m = \sqrt{k/m}$ is the (angular) resonance frequency of the oscillator and $Q_m$ is its mechanical $Q$-factor (see the plot of $|\chi(\omega)|$ in Fig.~\ref{Fig1}b for $Q_m=10$). Usually, accelerometers are operated below their fundamental resonance frequency $\omega_m$, where $\chi(\omega) \approx 1/\omega_m^2$ exhibits an almost flat frequency-response. This naturally leads to a tradeoff between resolution and bandwidth, since the large resonance frequency required for high-speed operation results in vanishingly small displacements.  As a result, the performance of the displacement sensor constitutes a central figure of merit of an accelerometer.

In a cavity optomechanical system, a mechanically compliant electromagnetic cavity is used to resonantly-enhance read out of mechanical motion~\cite{Braginsky1977} (canonically, the motion of the end mirror of a Fabry-Perot cavity). Such systems have enabled motion detection measurements with an imprecision at or below the standard quantum limit (SQL)~\cite{tittonen_interferometric_1999,Anetsberger_2010,Regal_2008}, corresponding to the position uncertainty in the quantum ground-state of the mechanical object.  Clever quantum back-action evading techniques~\cite{hertzberg_back-action-evading_2009} aside, only for an ideal cavity system (no parasitic losses) can the actual displacement sensitivity reach the SQL due to fluctuating radiation pressure forces arising from shot noise of the probe light~\cite{clerk_introduction_2010}.  The average radiation pressure force, on the otherhand, can be quite large in micro- and nano-scale optomechanical devices, and offers the unique capability to control the sensor bandwidth via the optical spring effect~\cite{Corbitt2007,lin_mechanical_2009} and the sensor's effective temperature via passive damping~\cite{kippenberg_cavity_2007} or feedback cold-damping~\cite{genes_ground-state_2008,kleckner_sub-kelvin_2006}.

In this work, we utilize an integrated silicon-nitride (SiN) \emph{zipper} photonic crystal optomechanical cavity~\cite{eichenfield_picogram-_2009} to provide shot-noise-limited read out of mechanical motion with imprecision at the SQL, enabling high-bandwidth and high-resolution acceleration sensing. The resolution of an accelerometer can be quantified by a noise-equivalent acceleration, $\mathrm{NEA} = \sqrt{a_\mathrm{th}^2 + a_\mathrm{det}^2 + a_\mathrm{add}^2}$ in units of $\mathrm{g/\sqrt{Hz}}$ ($1\ \mathrm{g} = 9.81\ \mathrm{m/s^2}$).  The first term in the NEA is due to thermal Brownian motion of the test mass (see appendix~\ref{SecThermalMotion})~\cite{yasumura_quality_2000} and is given by,
\begin{equation}\label{Eqath}
a_\mathrm{th} = \sqrt{\frac{4k_\mathrm{B}T\omega_m}{mQ_m}},
\end{equation}
while the remaining two terms arise from the aforementioned displacement readout noise ($a_\mathrm{det}$) and added noise (back-action) onto the test mass due to the act of measurement ($a_\mathrm{add}$, see appendix~\ref{SecBackactionNoise}).  Fundamental to minimizing the NEA is a reduction in the intrinsic thermal noise, $a_\mathrm{th}$, which according to equation~(\ref{Eqath}), requires one to maximize the mass-$Q$ product at a given $\omega_m$. In most commercial accelerometers, the $Q$-factor is relatively low, which demands large test masses for high resolution. In contrast, in the zipper cavity devices presented here, we use nano-tether suspension of a nano-gram test mass to yield high intrinsic mechanical $Q$-factors ($1-2\times 10^6$), and strong thermo-optomechanical back-action to damp and cool the thermal motion of the test mass.  
  
\begin{figure}[t]
\includegraphics[width=\columnwidth]{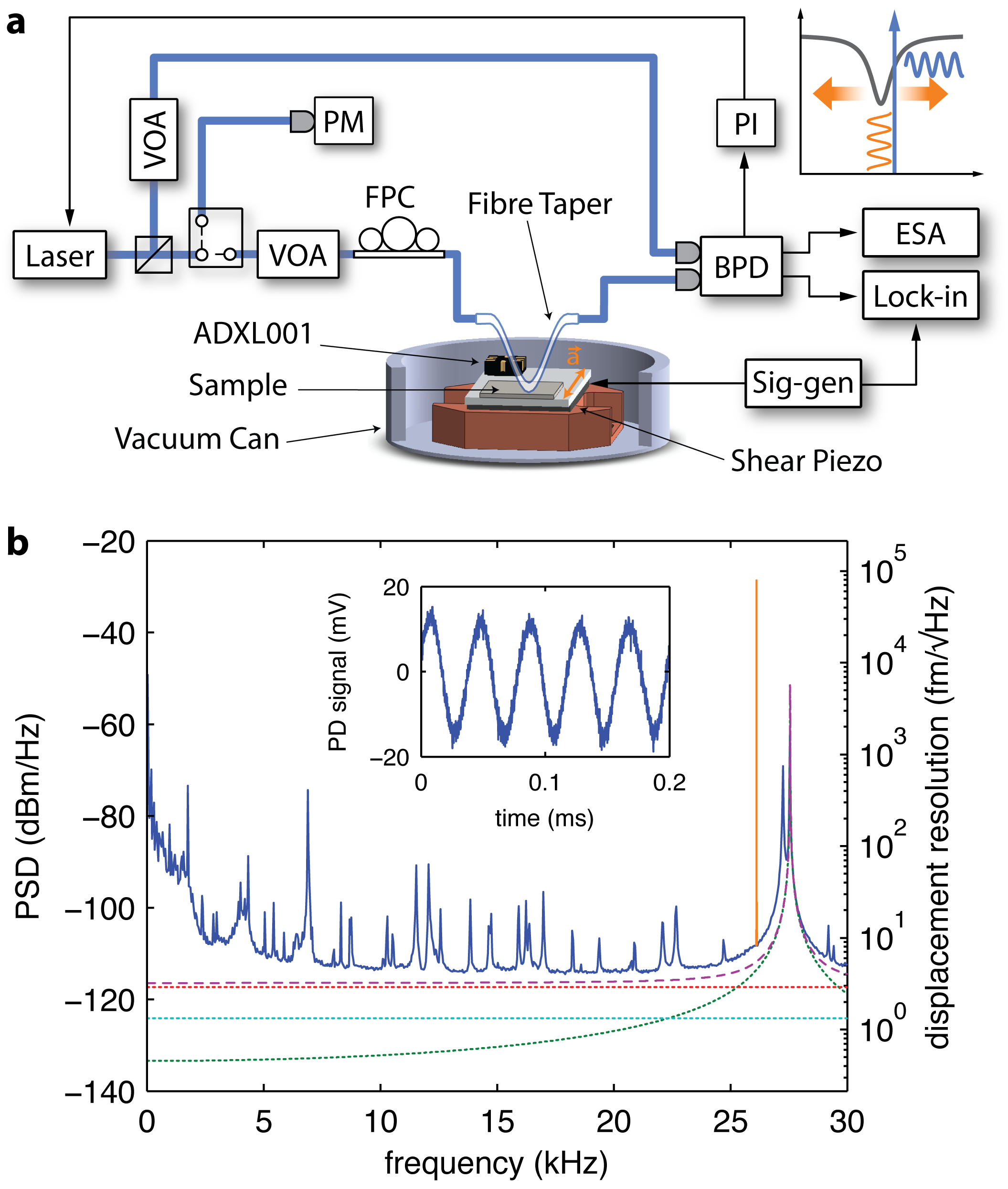}
\caption{\label{Fig2} \textbf{Experimental system and noise data.}
	\textbf{a}, Laser light used to probe the zipper cavity motion is split with a beamsplitter; the signal arm is sent through a fiber polarization controller (FPC) and a fiber taper, which is coupled to the optical cavity, while the other arm is sent directly to a balanced photo-detector (BPD).  Variable optical attenuators (VOA) in each arm balance the powers, and a power meter (PM) is used to calibrate the probe power.  The BPD signal is sent to a proportional-integral controller (PI) -- locking the laser half a linewidth red-detuned from the optical resonance. The sample is mounted on a shake table comprised of a shear piezo.  Transduced accelerations are measured using either an electronic spectrum analyzer (ESA) or a lock-in amplifier.
		\textbf{b}, The left axes show an optical power spectral density (PSD) plot of the BPD signal showing mechanical modes at 27.5~kHz (green).  The right axis shows the equivalent displacement noise.  The tone at 26~kHz (orange) is transduction of a tone applied to the shear piezo corresponding to an acceleration of 38.9~mg. The dashed and dotted lines are theoretical noise levels for shot noise (red), detector noise (cyan), thermal noise (green), and the total of all noise contributions (purple).  The inset is a time trace of the transduction of an applied acceleration of 35.6~mg at 25~kHz.}
\end{figure}

Figure~\ref{Fig1}c shows a scanning-electron microscope image of the device studied here, with the test mass structure and nano-tethers highlighted in green. The fundamental in-plane mechanical mode of this structure is depicted in Fig.~\ref{Fig1}e and is measured to have a frequency of $f_m = 27.5\ \mathrm{kHz}$, in good agreement with finite-element-method simulations from which we also extract a motional mass of $m=10\times 10^{-12}\ \mathrm{kg}$. The measured mechanical $Q$-factor is $Q_m = 1.4\times 10^6$ in vacuum (see appendix~\ref{SecMechQAcorr}), which results in an estimated $a_\mathrm{th} = 1.4\ \mathrm{\mu g/\sqrt{Hz}}$.  The region highlighted in pink corresponds to the zipper optical cavity used for monitoring test mass motion, a zoom-in of which can be seen in Figure~\ref{Fig1}d. The cavity consists of two patterned photonic crystal nanobeams, one attached to the test mass (bottom) and one anchored to the bulk (top). The device in Fig.~\ref{Fig1}c is designed to operate in the telecom band, with a measured optical mode resonance at $\lambda_o=1537$~nm and an optical $Q$-factor of $Q_o=9,500$.  With the optical cavity field being largely confined to the slot between the nanobeams, the optical resonance frequency is sensitively coupled to relative motion of the nanobeams in the plane of the device (the $\hat{x}$-direction in Fig.~\ref{Fig1}c). A displacement of the test mass caused by an in-plane acceleration of the supporting microchip can then be read-out optically using the setup shown in Fig.~\ref{Fig2}a, where the optical transmission through the photonic crystal cavity is monitored via an evanescently-coupled fiber taper waveguide~\cite{michael_optical_2007} anchored to the rigid side of the cavity.
Utilizing a narrow bandwidth ($<300$~kHz) laser source, with laser frequency detuned to the red side of the cavity resonance, fluctuations of the resonance frequency due to motion of the test mass are translated linearly into amplitude-fluctuations of the transmitted laser light field (see inset in Fig.~\ref{Fig2}a and appendix~\ref{SecOmTransd}). A balanced detection scheme allows for efficient rejection of laser amplitude noise, yielding shot-noise limited detection for frequencies above $\sim 1\ \mathrm{kHz}$.

Figure~\ref{Fig2}b shows the electronic power spectral density (PSD) of the optically transduced signal obtained from the device in Fig.~\ref{Fig1}c. The cavity was driven with an incident laser power of $P_\mathrm{in} = 116\ \mathrm{\mu W}$, yielding an intracavity photon-number of $\approx 430$. The two peaks around $27.5$~kHz arise from thermal Brownian motion of the fundamental in- and out-of-plane mechanical eigenmodes of the suspended test mass. The transduced signal level of the fundamental in-plane resonance, the mode used for acceleration sensing, is consistent with an optomechanical coupling constant of $g_\mathrm{OM} = 2\pi\times 5.5\ \mathrm{GHz/nm}$, where $g_\mathrm{OM}\equiv \partial\omega_o/\partial x$ is defined as the optical cavity frequency shift per unit displacement. The dotted green line depicts the theoretical thermal noise background of this mode. The series of sharp features between zero frequency (DC) and 15~kHz are due to mechanical resonances of the anchored fiber-taper. The noise background level of Fig.~\ref{Fig2}b is dominated by photon shot-noise, an estimate of which is indicated by the red dotted line. The cyan dotted line in Fig.~\ref{Fig2}b corresponds to the electronic photodetector noise, and the purple dashed line represents the sum of all noise terms. The broad noise at lower frequencies arises from fiber taper motion and acoustic pick-up from the environment. The right-hand axis in Fig.~\ref{Fig2}b quantifies the optically transduced PSD in units of an equivalent transduced displacement amplitude of the fundamental in-plane mode of the test mass, showing a measured shot-noise-dominated displacement imprecision of $4\ \mathrm{fm/\sqrt{Hz}}$ (the estimated on-resonance quantum-back-action displacement noise is $23\ \mathrm{fm/\sqrt{Hz}}$, and the corresponding on-resonance SQL is $2.8\ \mathrm{fm/\sqrt{Hz}}$; see appendix~\ref{SecBackactionNoise}).

At this optical power the observed linewidth of the mechanical mode is $\approx 2$~Hz, roughly 100 times larger than the low power linewidth.  As modeled in appendix~\ref{SecBackactTheory}, the measured mechanical damping is a result of radiation pressure dynamical back-action, enhanced by slow thermo-optical tuning of the cavity which provides the necessary phase-lag for efficient velocity damping.  Damping of the mechanical resonance is typically used to reduce the ringing transient response of the sensor when subjected to a shock input~\cite{y._t._li_air_1970}. In contrast to conventional gas-damping employed in MEMS sensors~\cite{allen_accelerometer_1989}, optomechanical back-action damping also cools the mechanical resonator~\cite{genes_ground-state_2008}.  The measured effective temperature of the fundamental in-plane mode of the test mass, as determined from the area under the $27$~kHz resonance line in Fig.~\ref{Fig2}b, is $T_\mathrm{eff} \approx 3\ \mathrm{K}$.  This combination of damping and cooling keeps the ratio of $T_\mathrm{eff}/Q_m$ fixed, and does not degrade the thermally-limited acceleration resolution of the sensor.

In order to carefully calibrate the accelerometric performance of the device, the sample is mounted onto a shake table driven by a shear piezo actuator (see appendix~\ref{SecShakeTableDescr}). Applying a sinusoidal voltage to the piezo results in a harmonic acceleration $a(\omega)$, and thereby a modulation of the transmitted optical power. The optical power in the modulation sideband is given by (see appendix~\ref{SecOmTransd})
\begin{equation}
P_m(\omega) = \left( 1-T_d \right) \frac{Q_o}{\omega_o} P_\mathrm{in} \, g_\mathrm{OM} \, \left| \chi(\omega) \, a(\omega)  \right|,
\end{equation}
where $Q_o$ is the optical $Q$-factor ($=9,500$), $\omega_o$ is the optical resonance frequency, $T_d$ is the relative cavity transmission on resonance ($=0.88$), and the laser is half a linewidth detuned.
The narrow tone at $26$~kHz in Fig.~\ref{Fig2}b (orange) arises from an applied rms-acceleration of $a_\mathrm{rms} = 38.9\ \mathrm{mg}$, calibrated using two commercial accelerometers mounted on the shake table (see appendix~\ref{SecShakeTableDescr}). From the signal-to-noise-ratio of this calibration tone we estimate $a_\mathrm{th} = 2.0\ \mathrm{\mu g/\sqrt{Hz}}$, comparable to the theoretical value of $a_\mathrm{th} = 1.4\ \mathrm{\mu g/\sqrt{Hz}}$.  For a driving tone at $10$~kHz, we measure $a_\mathrm{min} \approx 10\ \mathrm{\mu g/\sqrt{Hz}}$, limited in this case by photon shot noise.  The dynamic range over which the sensor is linear at a drive frequency of 10~kHz has also been measured (see appendix~\ref{SecLinDynRange}), and is found to be $> 49$~dB (up to $\sim 10$~g accelerations, limited by the maximum output voltage of the piezo shaker drive electronics).
 
\begin{figure*}[t]
\includegraphics[width=2\columnwidth]{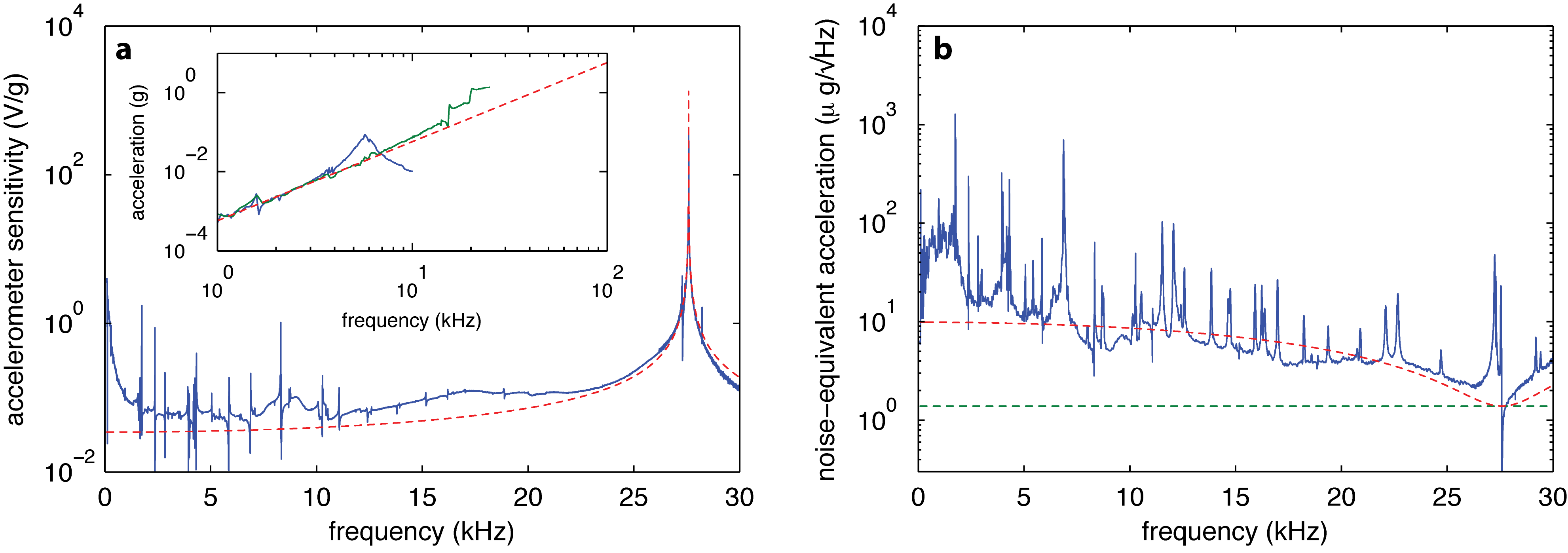}
\caption{\label{Fig3} \textbf{Frequency-dependence of sensitivity and resolution.}
	\textbf{a}, Sensitivity curve as function of frequency, obtained by driving the shear piezo with a sinusoidal voltage and measuring the amplitude of the resulting voltage modulation of the BPD signal using a lock-in amplifier. The dashed red line corresponds to the theoretical expectation for the sensitivity without fit parameters. The inset shows data from commercial accelerometers also attached to the shake table (blue and green curves), which are used for calibrating the applied acceleration.
	\textbf{b}, Frequency-dependent noise-equivalent acceleration (NEA) of the device quantifying its broadband-resolution. The plot is obtained by taking the PSD in Fig.~\ref{Fig2}b and normalising it by the sensitivity curve in \textbf{a}. The dashed red line depicts the theoretical expectation for the NEA given shot-noise and thermal noise limitations. The green dashed curve corresponds to the thermal noise ($a_\mathrm{th}$).}
\end{figure*}

Figure~\ref{Fig3}a shows the demodulated photodiode signal normalized to the applied acceleration as a function of drive frequency, corresponding to the frequency dependent acceleration sensitivity of the zipper cavity (the inset of Fig.~\ref{Fig3}a shows data from the commercial accelerometers used to calibrate the applied acceleration). The dashed red line is the theoretical calculation of the sensitivity without fit parameters and shows excellent agreement. The sharp Fano-shaped features for lower frequencies can again be attributed to mechanical resonances of the fiber-taper waveguide. The broad region of apparent higher-sensitivity around $15$~kHz is due to an underestimate of the applied acceleration arising from an acoustic resonance of the shake table.

The calibrated frequency-dependent NEA, shown in Fig.~\ref{Fig3}b, is obtained by normalizing the ESA noise spectrum (Fig.~\ref{Fig2}b) by the sensitivity curve (Fig.~\ref{Fig3}a). Between 25--30~kHz the resolution is limited by the thermal noise of the oscillator, while from 5--25~kHz shot-noise limits the resolution to $\approx 10\ \mathrm{\mu g/\sqrt{Hz}}$. For frequencies lower than $5$~kHz, motion of the fiber-taper waveguide and the environment contribute extra noise. The sharp Fano-shaped feature at $27$~kHz arises from interference with the fundamental out-of-plane mode of the test mass. The dashed red curve corresponds to a theoretical estimate of the NEA which shows good agreement. The dashed green line is the fundamental thermal sensing limit. 

\begin{figure*}[t]
\includegraphics[width=2\columnwidth]{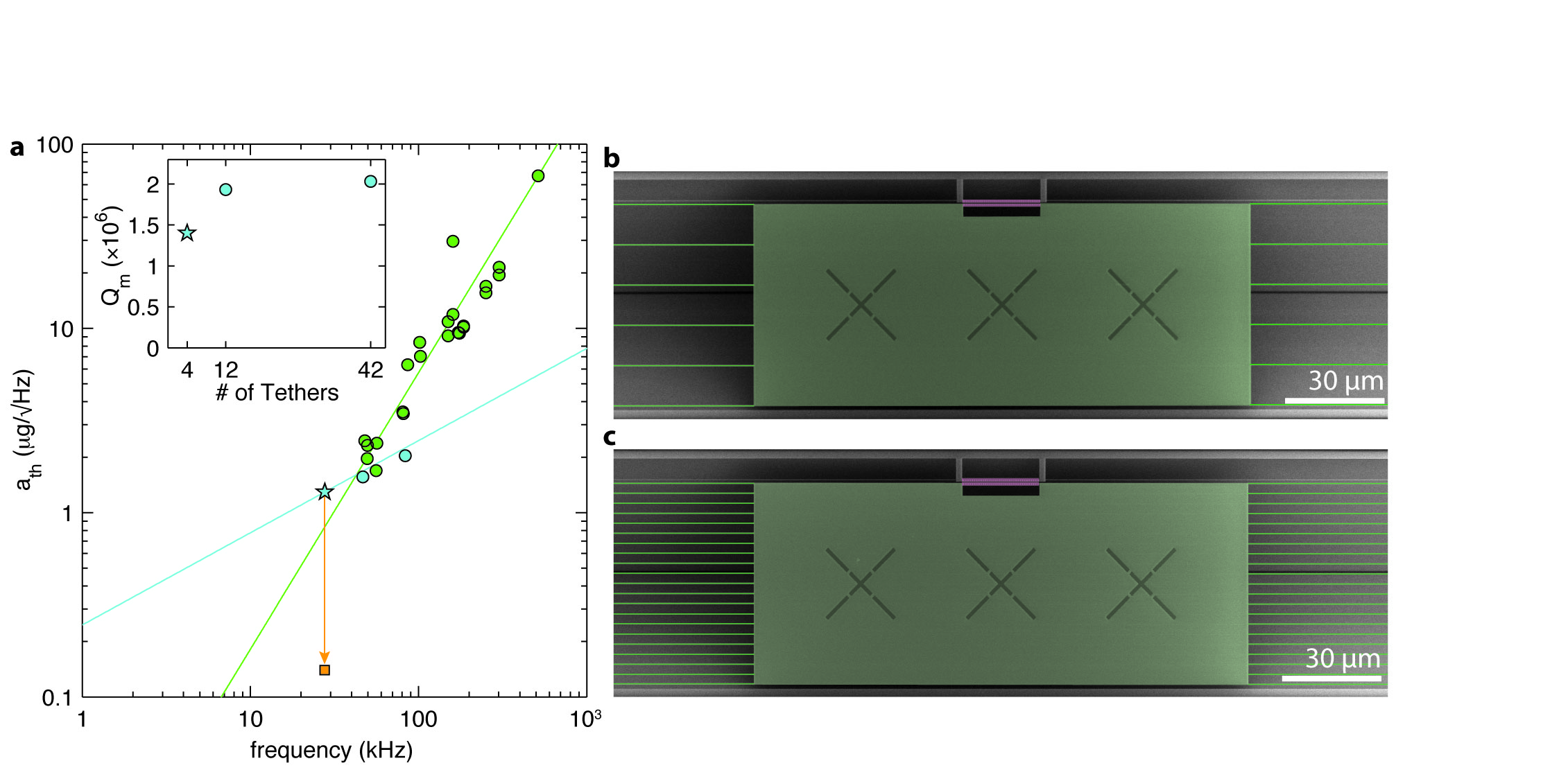}
\caption{\label{Fig4} \textbf{Independent tuning of bandwidth and resolution.}
	\textbf{a}, Thermal acceleration noise density of measured devices (green and cyan data points). The starred device is presented in the text. The green line is that traversed for adding mass with fixed $Q_{m}$ and spring constant $k$, while the cyan line is obtained for varying $k$ while keeping $Q_{m}$ and $m$ fixed. Varying both $m$ and $k$ allows for independent control of bandwidth and resolution, for example along the orange line, where $k/m$ is constant. The orange square represents theoretical device performance for 100 times increased test mass as compared to that in the text. The inset shows $Q_{m}$ for the devices corresponding to the cyan circles in \textbf{a} versus number of  nano-tethers attached to the test mass.  
	\textbf{b--c}, False-color SEM images of devices with 12 (42) nano-tethers and $f_{m}=46\ \mathrm{kHz}$ ($83\ \mathrm{kHz}$).}
\end{figure*}

The device platform demonstrated here straightforwardly allows for further reduction of the NEA. For instance, $a_\mathrm{th}$ can be reduced further by increasing the test mass $m$. In a preliminary study, we have fabricated a series of devices with test masses ranging from $100\times 10^{-15}\ \mathrm{kg}$ to $35\times10^{-12}\ \mathrm{kg}$ and recorded their mechanical frequency and $Q$-factor. Figure~\ref{Fig4}a depicts the calculated $a_\mathrm{th}$ versus the mechanical frequency of the studied devices, which roughly scales with $a_\mathrm{th} \propto \omega_m^{3/2}$ (green line).  Adding mass alone also results in a reduction of the sensor bandwidth; however, by scaling the number of nano-tether suspensions with the test mass size (see Figures~\ref{Fig4}b and c) the bandwidth can be kept constant. Moreover, as shown in the inset of Fig.~\ref{Fig4}a, we have found that adding nano-tethers does not result in a degradation of the mechanical $Q$-factor. Simultaneously scaling the width of the test mass and the number of nano-tethers by a factor of $100$ from the device shown in Fig.~\ref{Fig1}c, to a mass of $m = 10^{-9}\ \mathrm{kg}$, should reduce the thermal NEA to $\sim 150\ \mathrm{ng/\sqrt{Hz}}$ while maintaining a sensor bandwidth of $25$~kHz. Critically, for $g_\mathrm{OM} = 2\pi\times 100\ \mathrm{GHz/nm}$ as measured in previous zipper cavity structures~\cite{eichenfield_picogram-_2009}, the optical input power required to reach this resolution across the entire sensor bandwidth is still sub-milliwatt ($\sim 850\ \mathrm{\mu W}$).

With a demonstrated acceleration resolution on the order of a few $\mathrm{\mu g/\sqrt{Hz}}$ and a bandwidth above $25$~kHz, the zipper cavity device presented here shows performance metrics orders of magnitude better than other optical accelerometers~\cite{krishnamoorthy_-plane_2008,zandi_-plane_2010} and comparable to the best commercial sensors~\cite{Q-Flex_datasheet}.  These devices, formed from a silicon chip, also allow for the integration of electrostatic tuning-capacitors~\cite{winger_chip-scale_2011}, fiber-coupled on-chip waveguides~\cite{noell_applications_2002}, and on-chip electronics, all of which enables convenient, small form-factor packaging, and eliminates the need for expensive tunable lasers.  In addition, nanoscale optomechanical cavities such as the zipper cavity studied here, offer the unique resource of strong radiation-pressure back-action. The optical spring effect, for example, allows for dynamic tuning of the mechanical resonance frequency, which can increase the low-frequency displacement response (inverse quadratically with frequency) and decrease thermal noise (with the square root of frequency). Similar zipper cavity devices have shown low power (sub-mW) optical tuning of the mechanical resonance frequency over 10's of MHz ($>200 \%$ of $\omega_m$) into a regime where the mechanical structure is almost entirely suspended by the optical field~\cite{eichenfield_picogram-_2009}.   Also, as demonstrated here, back-action cooling provides a resource to damp the response of the oscillator without compromising the resolution.  Combining all of these attributes should allow not only for a new class of chip-scale accelerometers, but other precision displacement-based sensors of, for example, mass, force, and rotation.

\noindent\textbf{Acknowledgements} This work was supported by the DARPA QuASaR program through a grant from ARO. T.D.B.\ acknowledges support from the NSF GRFP under grant number 0703267.

\onecolumngrid

\clearpage


\appendix

\renewcommand{\thefigure}{A\arabic{figure}}
\renewcommand{\theequation}{A\arabic{equation}}
\setcounter{figure}{0}
\setcounter{equation}{0}
\makeatletter
	\@removefromreset{equation}{section}
\makeatother

\section*{Appendix}

\section{Oscillator susceptibility}

The oscillator susceptibility $\chi(\omega)$ given above follows from the differential equation of the harmonic oscillator:
\begin{equation}
m\ddot{x} + m\gamma \dot{x} + m\omega_m^2 x = F_\mathrm{appl}.
\end{equation}
Transforming to Fourier space, this reads
\begin{equation}
-\omega^2 x + i\omega \gamma x + \omega_m^2 x = \frac{F_\mathrm{appl}(\omega)}{m}.
\end{equation}
With $F_\mathrm{appl}(\omega)/m = a_\mathrm{appl}$, this yields the accelerometer response
\begin{eqnarray}
x(\omega) & = & \chi(\omega) a_\mathrm{appl}(\omega) \nonumber \\
 & = & \frac{1}{\omega_m^2-\omega^2+i\frac{\omega\omega_m}{Q_m}} a_\mathrm{appl}(\omega).
 \end{eqnarray}
This function has the following properties:
\begin{eqnarray}
\chi(0) & = & \frac{1}{\omega_m^2} = \frac{m}{k}, \\
\chi(\omega_m) & = & -i \frac{Q_m}{\omega_m^2} = -i Q_m \chi(0), \\
\chi(\omega \gg \omega_m) & \propto & \frac{1}{\omega^2}.
\end{eqnarray}
For the device studied here with $\omega_m = 27.5\ \mathrm{kHz}$, this gives an acceleration sensitivity of $\chi(0) = 329\ \mathrm{pm/g}$ with $g = 9.81\ \mathrm{m/s^2}$.

\section{Sample fabrication and design}

The presented accelerometer structures are defined in a 400~nm thick silicon nitride (SiN) layer formed on top of a $500\ \mathrm{\mu m}$ thick single-crystal silicon wafer. The SiN is stiochiometric and is grown in LPCVD under conditions that allow for large internal tensile stress ($\sigma = 800\ \mathrm{MPa}$). The accelerometer structures comprising the test mass, the support nano-tethers, and the zipper cavity are defined in a single electron-beam lithography step. The mask is transferred into the SiN layer using ICP/RIE dry-etching in a $\mathrm{SF_6/C_4F_8}$ plasma. Resist residues are removed in a combination of heated Microposit 1165 remover and Piranha solution (3:1 $\mathrm{H_2SO_4:H_2O_2}$) at $120^\circ\ \mathrm{C}$. The structures are undercut by anisotropic wet-etching in $70^\circ\ \mathrm{C}$ hot KOH and cleaned in a second Piranha etching step. Critical point dying in $\mathrm{CO}_2$ avoids collapsing of the zipper cavities. \\
The optical and mechanical structures are designed using finite-elements simulations performed in COMSOL Multiphysics (http://www.comsol.com/).

\section{Optical spectroscopy}

The sample is optically coupled via a near-field probe consisting of a tapered optical fiber.  The tapered fiber is brought in optical contact with the device using attocube nanopositioners. Aligned in parallel to the zipper nano-beams, the fiber taper is mechanically anchored on the struts attached to the rigid side of the zipper cavity. Launching light from a NewFocus Velocity tunable external-cavity diode laser into the fiber taper and monitoring the taper transmission then allows us to do resonant coherent spectroscopy of the cavity mode. Technical amplitude noise of the laser ($\sim 10\ \mathrm{dB}$ above the shot-noise level) is suppressed by a balanced detection scheme using a Newport 2117 balanced photodetector that features $\sim 20\ \mathrm{dB}$ common-mode noise rejection.

\section{Transmission function of side-coupled open cavity}

\begin{figure}
\includegraphics{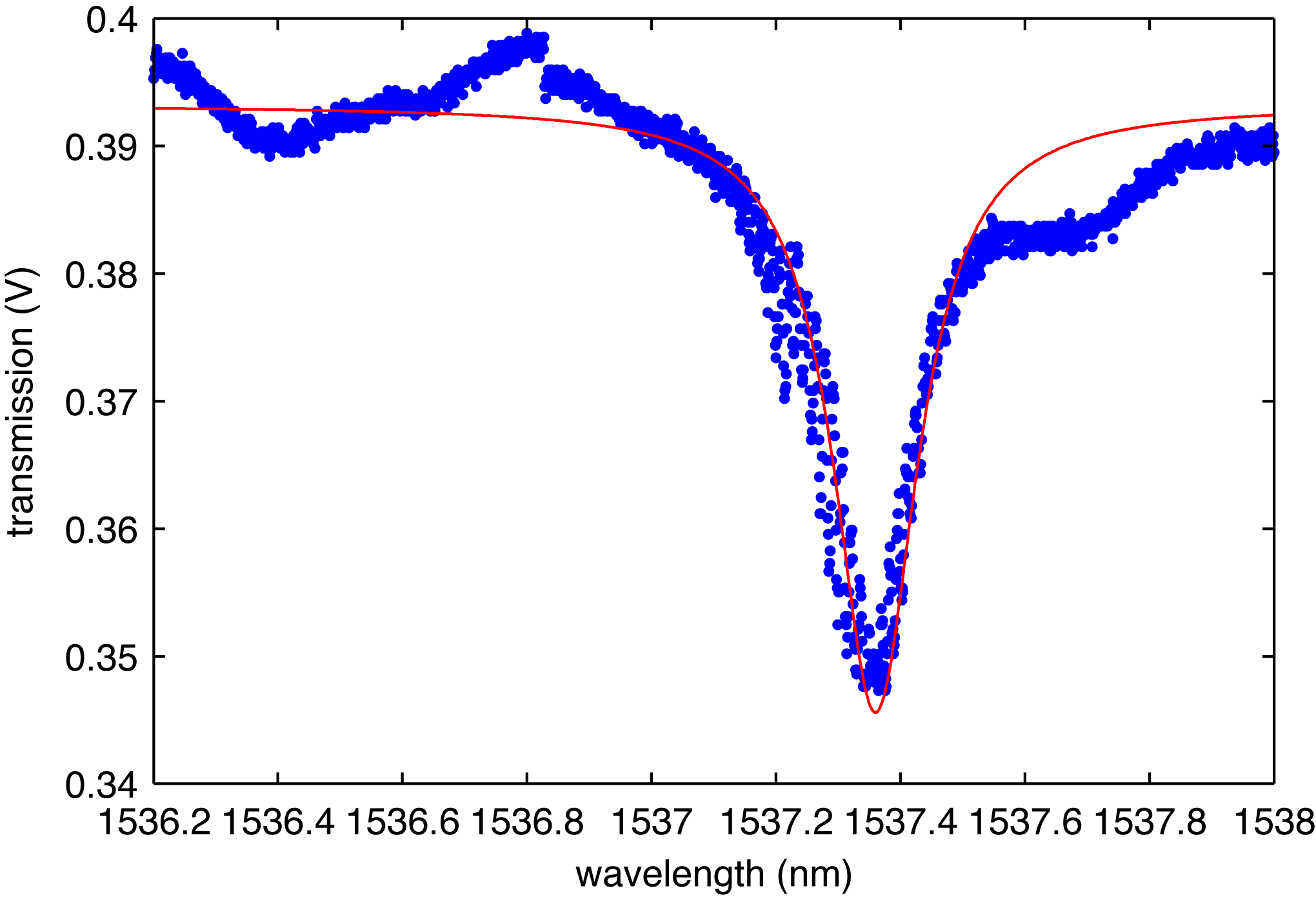}
\caption{\label{FigTransmCurve} \textbf{Example transmission curve of the zipper cavity.} The curve is obtained by scanning an external cavity diode laser across the cavity resonance at $\lambda_0 = 1537.36\ \mathrm{nm}$ while monitoring the fiber taper transmission. The dip exhibits an optical $Q$-factor of $Q_o = 9,\!500$ and a transmission dip on resonance of $T_d = 0.88$.}
\end{figure}

In order to calculate the intensity transmission profile $T(\omega)$ of a photonic-crystal resonator side-coupled by a fiber-taper waveguide, we start from the equation of motion of $\hat{a}$, the annihilation operator of the cavity field:
\begin{equation}\label{a_eq_mot}
\frac{\mathrm{d}\hat{a}}{\mathrm{d}t} = -\left( i\Delta + \frac{\kappa}{2} \right) \hat{a} + \sqrt{\frac{\kappa_e}{2}} \hat{a}_\mathrm{in} +  \sqrt{\kappa_i} \hat{a}_\mathrm{i} + \sqrt{\frac{\kappa_e}{2}} \hat{a}_\mathrm{-}.
\end{equation}
Here, $\Delta = \omega_\mathrm{l}  - \omega_\mathrm{c}$ is the laser-cavity detuning, $\kappa_e$ is the total taper-cavity coupling rate, $\kappa = \kappa_i + \kappa_e$ is the total cavity decay rate, with $\kappa_i$ the intrinsic cavity damping rate, and $\hat{a}_\mathrm{in}$ is the taper input field, which together with the output field $\hat{a}_\mathrm{out}$ obeys the boundary condition
\begin{equation}\label{EqBcond}
\hat{a}_\mathrm{in} + \hat{a}_\mathrm{out} = \sqrt{\frac{\kappa_e}{2}} \hat{a}.
\end{equation}
The last two terms on the right-hand-side of eq. (\ref{a_eq_mot}) represent the vacuum inputs due to coupling with the intrinsic (loss) bath of the cavity and the backward fiber taper waveguide mode, respectively (these input terms are ignored going forward as they are in the vacuum state and do not modify the classical field equations).  In steady state, where $\frac{\mathrm{d}\hat{a}}{\mathrm{d}t} \equiv 0$, the intracavity field operator is
\begin{equation}\label{EqA0Ss}
\hat{a}_0 = \sqrt{\frac{\kappa_e}{2}} \frac{\hat{a}_\mathrm{in}}{i\Delta + \frac{\kappa}{2}}.
\end{equation}
$\hat{a}_\mathrm{in}$ is normalized to the power  incident on the cavity $P_\mathrm{in}$ as $P_\mathrm{in} = \hbar \omega_\mathrm{l} \langle \hat{a}^\dagger_\mathrm{in} \hat{a}_\mathrm{in} \rangle$ such that the intracavity photon number is
\begin{equation}\label{Eqncav}
n_\mathrm{cav} = \langle \hat{a}^\dagger \hat{a} \rangle = \frac{\kappa_e}{2} \frac{1}{\Delta^2 + \frac{\kappa^2}{4}} \frac{P_\mathrm{in}}{\hbar\omega_\mathrm{l}}.
\end{equation}
Combining eq.~(\ref{EqBcond}) with eq.~(\ref{EqA0Ss}) yields the intensity transmission function 
\begin{equation}
T(\Delta) = \frac{|a_\mathrm{out}|^2}{|a_\mathrm{in}|^2} = 1 - \frac{\kappa_e}{4} \frac{2\kappa - \kappa_e}{\Delta^2 + \frac{\kappa^2}{4}}.
\end{equation}
This function describes a Lorentzian absorption curve that dips to $T_d = \frac{\kappa_i^2}{\kappa^2}$ at $\Delta = 0$. Figure~\ref{FigTransmCurve} shows an example transmission curve of the device studied in this work obtained by scanning an external cavity diode laser across the fundamental resonance of the zipper cavity.
The slope of the curve is given by 
\begin{equation}
\frac{\mathrm{d}T}{\mathrm{d}\Delta} = \frac{\kappa_e}{2} \frac{2\kappa-\kappa_e}{\left( \Delta^2 + \frac{\kappa^2}{4} \right)^2} \Delta.
\end{equation}
Usually, we lock the probe laser to a red-side detuning of $\Delta = -\kappa/2$, where the transduction is maximum for fixed $n_\mathrm{cav}$. At that detuning, the intracavity photon number is given by
\begin{equation} \label{Eqncav2}
n_\mathrm{cav,\kappa/2} = \left( 1-\sqrt{T_d} \right) \frac{Q_o}{\omega_o} \frac{P_\mathrm{in}}{\hbar\omega_l}
\end{equation}
and the slope of the transmission curve is
\begin{eqnarray}
\frac{\mathrm{d}T}{\mathrm{d}\Delta} \Big|_{\Delta = -\frac{\kappa}{2}} & = & -\frac{\kappa_e \left( 2\kappa-\kappa_e \right)}{\kappa^3} \\
 & = & -\left( 1- T_d \right) \frac{Q_o}{\omega_o}.\label{EqdTdDelta}
\end{eqnarray}

\section{Derivation of the optomechanical accelerometer transduction}\label{SecOmTransd}

Our device operates deep in the sideband unresolved regime, where $\omega_m \ll \kappa$ ($\omega_m = 2\pi\times 27.5\ \mathrm{kHz}$, $\kappa = \omega_c/Q_o = 129\ \mathrm{GHz}$).
In this regime, the intra-cavity field and hence the field transmitted through the cavity adiabatically follow changes in laser-cavity detuning $\Delta = \omega_l - \omega_c$ created by mechanical motion of the test-mass, $\Delta = g_\mathrm{OM} \, x$. In order to calculate the optical transmission change $\Delta T$ induced by a shift of the cavity resonance frequency $\Delta$, we can therefore approximate 
\begin{equation}
\Delta T = \frac{\mathrm{d}T}{\mathrm{d}\Delta} \Delta,
\end{equation}
such that the frequency component of the transmitted optical power arising from a displacement $x(\omega)$ is given by
\begin{equation}\label{Eqtransd}
P_\mathrm{m}(\omega) = \frac{\mathrm{d}T}{\mathrm{d}\Delta} \eta_\mathrm{in} P_\mathrm{in} g_\mathrm{OM}  \, x(\omega),
\end{equation}
where $P_\mathrm{in}$ is the input power in the fiber taper waveguide at the zipper cavity and $\eta_\mathrm{in}$ quantifies the optical loss in the fiber taper waveguide between the cavity and the detector via $\eta_\mathrm{in} = P_\mathrm{det}/P_\mathrm{in}$, where $P_\mathrm{det}$ is the optical power reaching the detector. This formula relates frequency components of the transmitted optical power modulation to the mechanical motion of the test-mass. With eq.~(\ref{Eqncav2}) and eq.~(\ref{EqdTdDelta}), this becomes
\begin{equation}\label{EqOMtransd}
P_\mathrm{m}(\Delta=\kappa/2) = \left( 1-T_d \right)\frac{Q_o}{\omega_o} g_\mathrm{OM} \, \eta_\mathrm{in} P_\mathrm{in} \, x.
\end{equation}

This optical power is measured on a Newport 2117 balanced photo-detector with switchable transimpedance gain (in these experiments we use $g_\mathrm{ti} = 49,\!600\ \mathrm{V/W}$), generating a voltage output of $V_\mathrm{m} = g_\mathrm{ti} P_\mathrm{m}$. An electronic spectrum analyzer (ESA) calculates the electrical power spectral density of this optical sideband in units of $V_\mathrm{m}^2/Z$ with $Z = 50\ \mathrm{\Omega}$ and expresses it in dBm/Hz. The conversion follows the relation
\begin{equation}\label{EqPSDESA}
\mathrm{PSD_{ESA}(\omega)} = 10 \cdot \log \left[ \frac{ \left( g_\mathrm{ti} P_\mathrm{m}(\omega) \right)^2}{Z} \cdot 1,\!000 \right].
\end{equation}

Careful calibration of the parameters in eq.~(\ref{EqOMtransd}) and eq.~(\ref{EqPSDESA}) as well as the optical input power, allows one to calculate the optomechanical coupling $g_\mathrm{OM}$ from the magnitude of the (known) thermal Brownian motion noise of the mechanical oscillator. In the measurements presented in Figs.~\ref{Fig2} and \ref{Fig3}, we have $T_d = 0.87$, $Q_o = 9,\!500$, $\omega_o = 2\pi \times 195\ \mathrm{THz}$, and $\eta_\mathrm{in} = 0.57$.  At low optical input power, where negligible back-action cooling is being performed on the fundamental in-plane mechanical mode of the suspended test mass and the mode's effective temperature is the temperature of the room temperature bath ($T \sim 300K$), the optomechanical coupling constant is estimated to be $g_\mathrm{OM} = 2\pi \times 5.5\ \mathrm{GHz/nm}$ from the area under the Lorentzian centered at $27$~kHz of the optically transduced displacement noise PSD.  This corresponds to an optical displacement sensitivity of $P_m/x = 3.7\ \mathrm{nW/pm}$ for the fundamental in-plane mechanical mode of the suspended test mass. From electromagnetic finite-elements simulations we calculate $g_\mathrm{OM} = 2\pi \times 13.5\ \mathrm{GHz/nm}$ for dimensions of the zipper cavity as measured with a scanning electron microscope, in good agreement with the measured value.

\section{Acceleration sensitivity measurement}\label{SecShakeTableDescr}
For applying AC accelerations to our device, we constructed a shake table comprising a sample holder plate glued on a shear piezo actuator. Applying a sinusoidal AC-voltage to the piezo creates a displacement $x_0 \sin(\omega_d t)$, which results in an applied acceleration of $-x_0 \omega_d^2 \sin(\omega_d t)$. For calibration of the shake table assembly, we use commercial accelerometers from Analog Devices of 5.5~kHz (ADXL103) and 22~kHz (ADXL001) bandwidth, respectively. In order to measure the frequency response of our optomechanical accelerometer, we apply a constant-voltage drive to the piezo and tune its frequency, while measuring the photodetector output on a lock-in amplifier. After normalizing for the $\omega_d$-dependence of the applied acceleration, this yields the frequency-dependent sensitivity of the device. Normalizing an optical noise PSD then allows us to calibrate the noise-floor of the accelerometer in terms of a noise-equivalent acceleration.

\section{Mechanical spectroscopy and autocorrelation method to determine mechanical quality factor}\label{SecMechQAcorr}

\begin{figure}[t]
\includegraphics{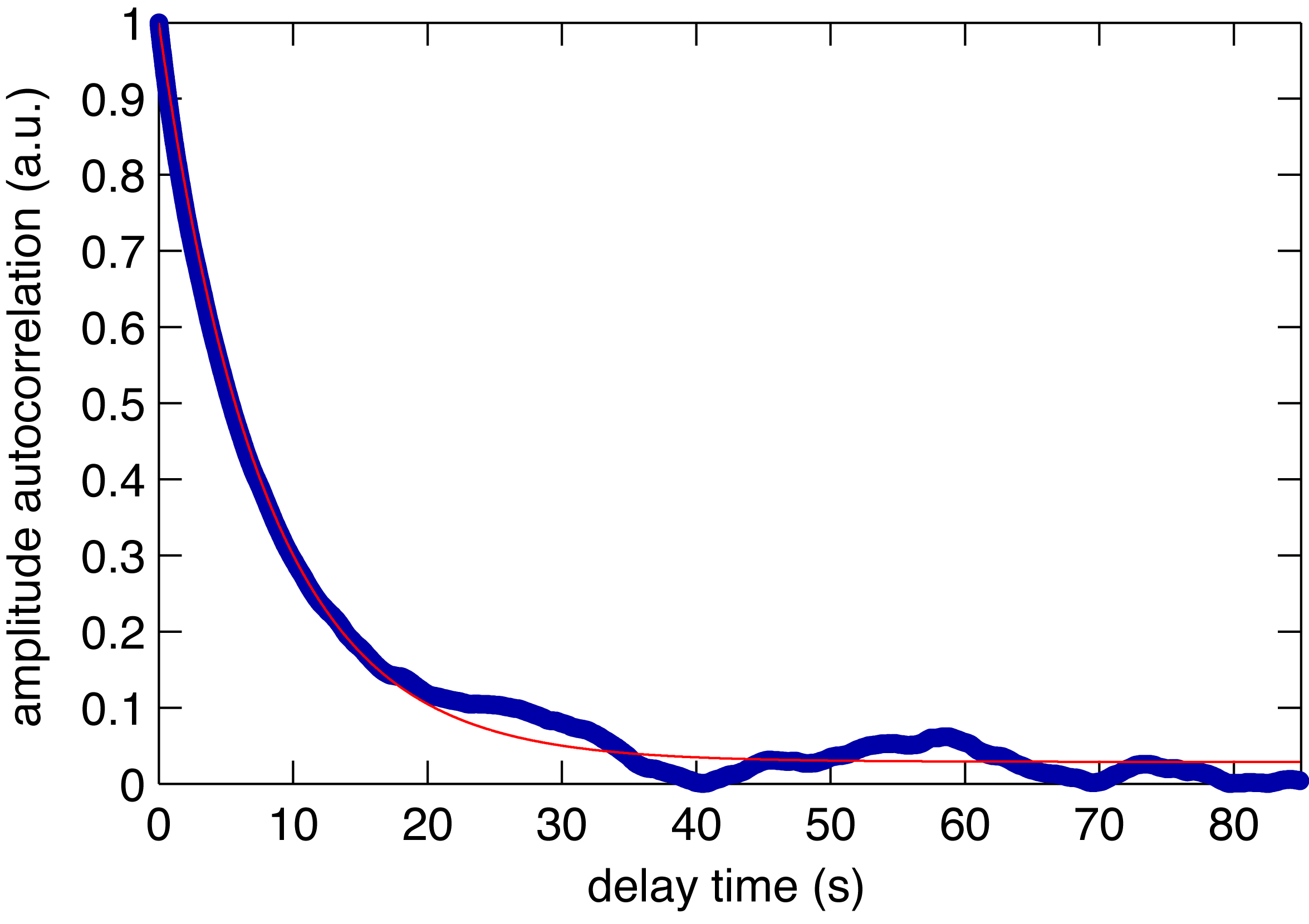}
\caption{\label{Fig_S_acorr} \textbf{Autocorrelation trace of the thermal noise driven mechanical amplitude.} The signal was obtained from computing the autocorrelation of the slowly varying magnitude of the mechanical motion returned from a lock-in amplifier. Fitting the trace with an exponential decay yields the time constant and thereby the mechanical $Q$-factor of the mode ($Q_m = 1.4 \times 10^6$).}
\end{figure}

Motion of the mechanical oscillator results in amplitude-modulation of laser light transmitted through the fiber taper which can be measured by monitoring the power spectral density of the detected balanced-photodiode photocurrent on an electronic spectrum analyzer (ESA) from which we can extract the resonance frequency and the total power in transduced sideband (proportional to mode temperature). However, the sub-Hz linewidths of our mechanical modes make establishing the quality factor from a measurement of the power spectral density on a spectrum analyzer infeasible because it requires a fractional stability of the frequency to greater than $\gg 1/Q_m \approx 5\times 10^{-7}$ over a period much longer than the decay time $Q_m/\omega_m \approx 10\ \mathrm{s}$. To overcome this limitation, we extract $Q_m$ from the autocorrelation function of the mechanical motion~\cite{stipe_noncontact_2001}. Since the system is driven by a Gaussian thermal noise process, the autocorrelation of the amplitude $\langle X(t)X(t+\tau) \rangle$ can be shown to decay as $\mathrm{e}^{-t/\tau}$ from which the quality factor can be obtained as $Q_m = \tau\omega_m$~\cite{stipe_noncontact_2001}. The slowly-varying envelope of $\langle X(t)\rangle$ is obtained from the magnitude channel of  a lock-in amplifier tuned to the mechanical resonance frequency with a bandwidth ($\approx 100\ \mathrm{Hz}$) much larger than the linewidth which ensures that small frequency diffusion does not affect the measurement of the envelope. To obtain the bare mechanical $Q$-factors the measurement is made at an optical power low enough to ensure there is no backaction. The autocorrelation is numerically computed and the decay is fit to an exponential curve with a constant (noise) offset. In Fig.~\ref{Fig_S_acorr} we show an autocorrelation trace of the device studied in this work calculated from $\approx 3000\ \mathrm{s}$ of data sampled at $100\ \mathrm{Hz}$ and fit it to find $\tau=7.85\ \mathrm{s}$ and for $\omega_m=2\pi\times27.5\ \mathrm{kHz}$ that $Q=1.4\times10^6$. For lower-$Q$ structures, it was confirmed that this technique agrees with a direct measurement of the linewidth from a spectrum analyzer. In order to avoid air-damping, measurements are carried out in vacuum.

\section{Optomechanical and thermo-optical backaction}\label{SecBackactTheory}

The relatively small test mass makes the device studied in this work highly susceptible to optomechanical and thermo-optical back-action effects. Such dispersive couplings are well known to renormalize the frequency and damping rate of the mechanical oscillator. In particular, thermo-optical coupling which arises from a refractive index change of the material upon the absorption of cavity photons plays a significant role in these devices due to the efficient thermal isolation of our nano-tethered test-masses in vacuum. Previous studies have shown strong modification of the optomechanical spring effect and damping in similar zipper cavity devices~\cite{eichenfield_picogram-_2009}.

The Supplementary Information of Ref.~\cite{eichenfield_picogram-_2009} gives a detailed derivation of the renormalized oscillator frequency and damping rate under the influence of optomechanical and thermo-optical coupling. The system of differential equations that describes the time evolution of the intra-cavity field $a$, the oscillator position $x$, and the cavity temperature shift $\Delta T$ is given by
\begin{eqnarray}
\dot{a} & = & -\left[ i\Delta - \left(g_\mathrm{OM} x + g_\mathrm{th} \Delta T \right) \right] a - \frac{\kappa}{2} a + \sqrt{\frac{\kappa_e}{2}} a_\mathrm{in} \\
\ddot{x} & = & -\gamma \dot{x} - \omega_m^2 x - \frac{\hbar g_\mathrm{OM}}{ m} \left| a \right|^2 \\
\dot{\Delta T} & = & -\gamma_\mathrm{th} \Delta T + \kappa_\mathrm{abs} c_\mathrm{th} \hbar \omega_c \left| a \right|^2,
\end{eqnarray}
where $g_\mathrm{th} = -(\mathrm{d}n/\mathrm{d}T)(\omega_c/n)$ is the thermo-optical tuning coefficient, $\mathrm{d}n/\mathrm{d}T$ is the thermo-optic coefficient of the material, $\kappa_\mathrm{abs}$ is the optical loss rate due to material absorption, $c_\mathrm{th}$ is the thermal heat capacity, and $\gamma_\mathrm{th}$ is the decay rate of the temperature. Linearizing these equations yields the static solutions
\begin{equation}
a_\mathrm{0} = \sqrt{\frac{\kappa_e}{2}} \frac{1}{i\Delta'+\kappa/2}, \quad
x_0 = \frac{\hbar g_\mathrm{OM}}{m\omega_m^2} \left| a_0 \right|^2, \quad
\Delta T_0 = \frac{\kappa_\mathrm{abs} }{ \gamma_\mathrm{th} } c_\mathrm{th} \hbar \omega_c \left| a_0 \right|^2
\end{equation}
with the renormalized detuning $\Delta' = \Delta - g_\mathrm{OM} x_0 - g_\mathrm{th} \Delta T_0$ arising from the static optomechanical and thermo-optical shift. Using a perturbation ansatz $x(t) = x_0 + \epsilon \cos(\omega_m t)$ one arrives after some algebraic manipulation at a modified harmonic oscillator equation for $x$ with a renormalized frequency $\omega_m'$ and damping rate $\gamma'$ given by
\begin{eqnarray}
\omega_m'^2 = \omega_m^2 & - & \frac{\hbar \omega_c n_\mathrm{cav} g_\mathrm{OM}^2 }{\omega_c m} \, \mathrm{Im} \left[ g(\omega_m) \right], \\
\gamma' = \gamma & + & \frac{\hbar \omega_c n_\mathrm{cav} g_\mathrm{OM}^2}{\omega_m \omega_c m} \, \mathrm{Re} \left[ g(\omega_m) \right], \label{EqOMTh_damping}
\end{eqnarray}
where the transfer function $g(\omega)$ is defined as
\begin{equation}
g(\omega) = f \frac{1+f'^*f^*}{|1+f'f|^2}
\end{equation}
with
\begin{equation}
f(\omega) = \frac{1}{i(\omega+\Delta')+\kappa/2} - \frac{1}{i(\omega-\Delta')+\kappa/2}
\end{equation}
and
\begin{equation}
f'(\omega) = -i \frac{ \Delta_\mathrm{th} \gamma_\mathrm{th} }{ i\omega + \gamma_\mathrm{th} },
\end{equation}
and $\Delta_\mathrm{th} = g_\mathrm{th} \Delta T_0$ is the static thermo-optical shift of the cavity resonance frequency. In the sideband unresolved regime where $\omega_m \ll \kappa$ and for thermal decay rates $\gamma_\mathrm{th}$ smaller than the mechanical frequency, an approximation of $g(\omega)$ yields
\begin{eqnarray}
\omega'^2 = \omega^2 & + & \frac{ 2\hbar n_\mathrm{cav} g_\mathrm{OM}^2 }{ m } \frac{ \Delta' }{ \Delta'^2 + \kappa^2/4 } \left[ \frac{ 1+W }{ 1+s } \right], \\
\gamma' = \gamma & + & \frac{ 2\hbar n_\mathrm{cav} g_\mathrm{OM}^2 }{ m } \frac{ \kappa \Delta' }{ (\Delta'^2 + \kappa^2/4)^2 } \left[ \frac{ 1+V }{ 1+s } \right],
\end{eqnarray}
with the correction factors
\begin{eqnarray}
W & = & -\left( \frac{ 2\Delta_\mathrm{th} }{ \kappa } \right) \left( \frac{ \gamma_\mathrm{th} }{ \omega_m } \right)^2 \left( \frac{ \kappa \Delta' }{ \Delta'^2 + \kappa^2/4 } \right), \\
V & = & \left( \frac{ 2\Delta_\mathrm{th} }{ \kappa } \right) \left( \frac{ \gamma_\mathrm{th} }{ \omega_m } \right)^2 \left( \frac{ \Delta' }{ \gamma_\mathrm{th} } \right) \\
\end{eqnarray}
and the saturation parameter
\begin{equation}
s = \left( \frac{ 2 \gamma_\mathrm{th} \Delta_\mathrm{th} \hbar \omega_c n_\mathrm{cav} }{ \omega_m } \frac{ \Delta' }{ \Delta'^2 + \kappa^2/4 } \right)^2 \left( 1 + \frac{ 1 }{ \Delta_\mathrm{th} } \left( \frac{ \Delta'^2 + \kappa^2/4 }{ \Delta'^2 } - \frac{ \omega_m^2 \kappa }{ \Delta' \gamma_\mathrm{th} } \right) \right)
\end{equation}

In the parameter regime of our devices, purely optomechanical back-action is a relatively weak effect due to the low optical $Q$-factor. For the parameters given above and for a pump laser with an incident power of $P_\mathrm{in} = 116\ \mathrm{\mu W}$ half a linewidth red-detuned from the cavity resonance, optomechanical back-action alone predicts a frequency shift of merely $\omega'_m - \omega_m = -2\pi \times 35.9\ \mathrm{Hz}$ and a damping factor of $\gamma'_m/\gamma  = 1.01$. 

\begin{figure}[t]
\includegraphics{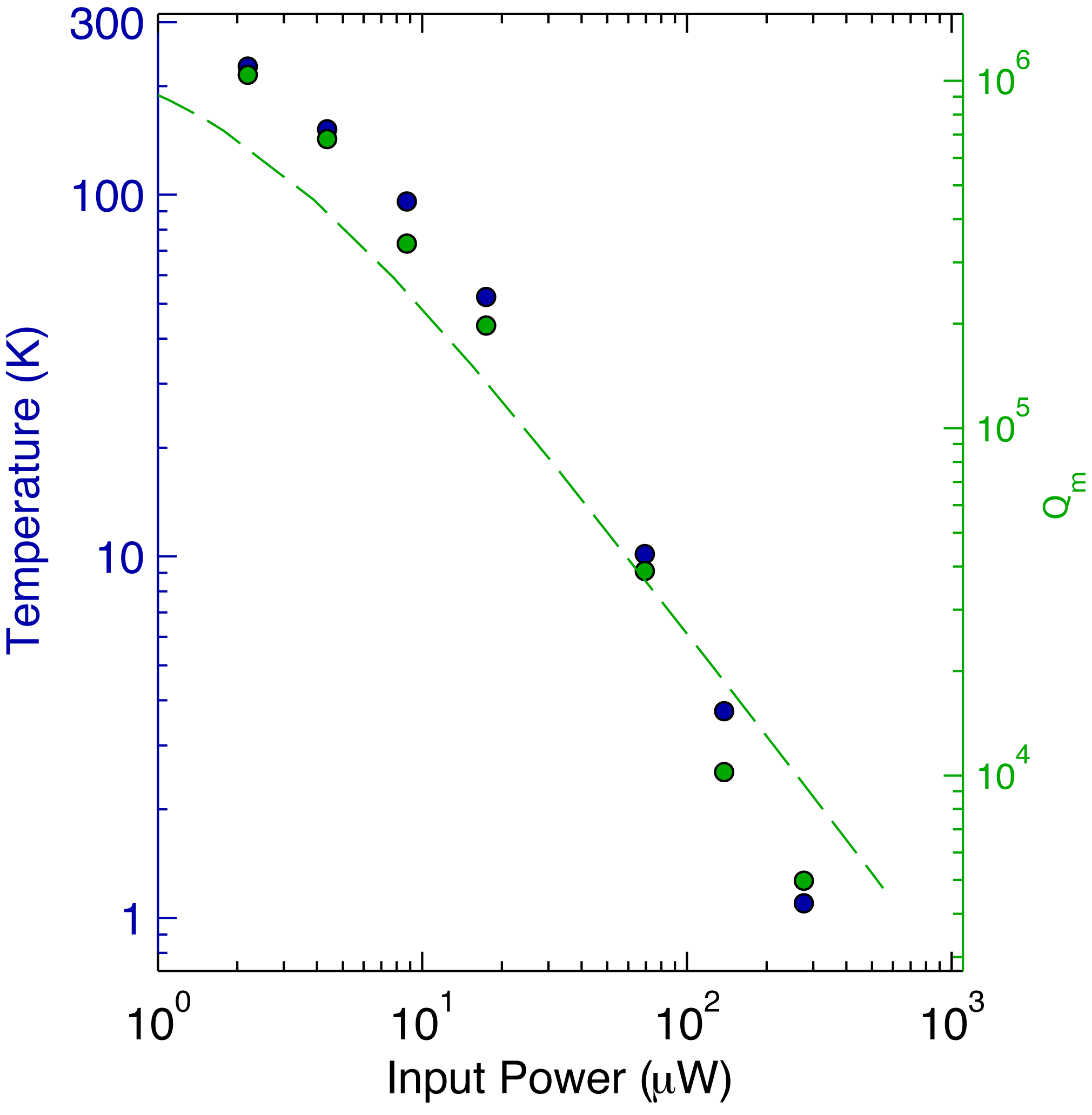}
\caption{\label{Fig_S_Cooling} \textbf{Demonstration of thermo-optomechanical damping and cooling.} The green bullets show measured $Q$-factors of the mechanical mode as function of the optical power, yielding thermo-optomechanical damping by a factor of $\approx 280$. The blue bullets show the corresponding optical power in the sideband generated by mechanical motion, proportional to the effective mode temperature. We observe cooling to $T_\mathrm{eff} \approx 1 \ \mathrm{K}$. The dashed green curve corresponds to a theoretical model that includes optomechanical and thermo-optical back-action.}
\end{figure}
In order to study the influence of thermo-optical back-action, we measured the $Q$-factor of the mechanical mode as function of the optical power launched into the cavity, shown as the green bullets in Fig.~\ref{Fig_S_Cooling}. When increasing the optical power to $P_\mathrm{in} \approx 300\ \mathrm{\mu W}$, which corresponds to an intracavity photon number of $n_\mathrm{cav} \approx 1\!,100 $, the $Q$-factor shows strong damping and is reduced by a factor of $\approx 200$. Similarly, we measure the area of the mechanical resonance peak from the optically transduced thermal noise PSD for a series of optical powers, and plot the inferred effective mode temperature as blue bullets in Fig.~\ref{Fig_S_Cooling}.  Clear in Fig.~\ref{Fig_S_Cooling} is that the effective mode temperature is dropping with the measured mechanical $Q$-factor.

The observed mechanical damping is much larger than the value predicted by pure optomechanical back-action and can be explained when including thermo-optical tuning. The green line in Fig.~\ref{Fig_S_Cooling} was obtained by calculating the modified $Q$-factor $Q'_m = \omega_m/\gamma'_m$ using eq.~(\ref{EqOMTh_damping}) with $\Delta_\mathrm{th} = -0.05 \kappa$ and $\gamma_\mathrm{th} = 2\pi \times 9.2\ \mathrm{kHz}$. The latter value is in good agreement with the one from \cite{eichenfield_picogram-_2009} ($\gamma_\mathrm{th} = 2\pi \times 10\ \mathrm{kHz}$), which suggests that the time constant of thermo-optical tuning is dominated by heat-flow from the zipper cavity region to the reservoir formed by the test-mass (or the bulk in the case of \cite{eichenfield_picogram-_2009}, respectively).

The obtained values for $\Delta_\mathrm{th}$ and $\gamma_\mathrm{th}$ result in correction factors of $V = 12,\!400$, $W = -0.011$, and a saturation parameter of $s \approx 3\times 10^{-36}$. Accordingly, we expect a significant thermo-optical correction to damping, as observed, but only a minor modification of the optomechanical spring: $\omega'-\omega = 2\pi \times 36.2\ \mathrm{Hz}$ for the pump power used in the experiment. Indeed, we observed a frequency shift of $101\ \mathrm{Hz}$, in reasonable agreement with the theoretical value.

\section{Analysis of optical noise power spectral densities}

As discussed above, noise power-spectral-densities (PSDs), such as those shown in Fig.~\ref{Fig2}b, arise from the contributions of various noise sources. In the following we derive expressions for their magnitudes.  Throughout the analysis below we work with single-sided PSDs, unless otherwise stated, as these are the PSDs measured in our experiment.

\subsection{Noise from thermal Brownian motion}\label{SecThermalMotion}
In contact with a heat-bath at room temperature, the test-mass oscillator is subjected to thermal Brownian motion. From the equipartition theorem, the root-mean-square displacement of a harmonic oscillator is given by
\begin{equation}
x_\mathrm{rms} = \sqrt{\frac{k_\mathrm{B}T}{k}}.
\end{equation}
If we assume the acceleration-noise exerted by the bath to be white, i.e.\ frequency-independent, its power-spectral density $S_{aa}^{th}$ has to obey
\begin{equation}
x_\mathrm{rms}^2 = \int_{0}^\infty  \left| \chi(\omega) \right|^2 S_{aa}^{th} \, \mathrm{d}\omega,
\end{equation}
such that thermal test-mass motion corresponds to a noise-equivalent acceleration (NEA) of
\begin{equation}\label{Eqath_SI}
a_\mathrm{th} = \sqrt{S_{aa}^{th}} = \sqrt{\frac{4k_BT\omega_m}{mQ_m}} = \sqrt{ \frac{ 4k_BT\gamma_m }{ m } }.
\end{equation}
In the device presented in this work, we have $\omega_m = 2\pi \times 27.5\ \mathrm{kHz}$, $m = 10^{-11}\ \mathrm{kg}$, $Q_m = 1.4\times 10^6$, $T = 295\ \mathrm{K}$, and therefore $a_\mathrm{th} = 1.4\ \mathrm{\mu g/\sqrt{Hz}}$. For a mass-on-a-spring oscillator with $\omega_m = \sqrt{k/m}$ this corresponds to
\begin{equation}
a_\mathrm{th} = \sqrt{\frac{4k_BT}{Q_m}} \frac{k^{^1/_4}}{m^{^3/_4}}.
\end{equation}
Driving the harmonic oscillator with susceptibility $\chi(\omega)$, this NEA translates into frequency-dependent displacement noise according to
\begin{equation}
S_{xx}^{th}(\omega) = \frac{4k_BT\omega_m}{mQ_m} \frac{1}{\left( \omega^2 - \omega_m^2 \right)^2 + \left(\frac{\omega \omega_m}{Q_m}\right)^2}.
\end{equation}
According to eq.~(\ref{Eqtransd}), the optical signal transduced by the cavity then exhibits a noise power-spectral density of
\begin{eqnarray}
S_{PP}^{th}(\omega) & = & \left| \frac{\mathrm{d}T}{\mathrm{d}\Delta} \right|^2 \eta_\mathrm{in}^2 P_\mathrm{in}^2 g_\mathrm{OM}^2 S_{xx}^{th} \\
 & = & \left( 1-T_d \right)^2 \frac{Q^2}{\omega_0^2} \eta_\mathrm{in}^2 P_\mathrm{in}^2 g_\mathrm{OM}^2 \frac{4k_BT\omega_m}{mQ_m} \frac{1}{(\omega^2-\omega_m^2)^2+\left( \frac{\omega\omega_m}{Q_m} \right)^2}.
\end{eqnarray}

Under the influence of thermo-optomechanical back-action discussed in appendix~\ref{SecBackactTheory}, the dynamic parameters $\omega_m$ and $Q_m$ have to be replaced by the renormalized values $\omega'_m$ and $Q'_m$. With the parameter values realized in this experiment, the optical noise arising from thermal Brownian motion corresponds to $\sqrt{S_{PP}^{th}(0)} = 0.96\ \mathrm{pW/\sqrt{Hz}}$.

\subsection{Optical shot noise}
Photon shot noise arises from the quantum nature of light and from the destructive character of optical measurements using photodiodes. The single-sided shot-noise power-spectral-density for light of frequency $\omega_0$ and power $P_\mathrm{det}$ incident on a photo-detector is frequency-independent and given by
\begin{equation}
S_{PP}^{SN} = \frac{2\hbar\omega_o P_\mathrm{det}}{\eta_\mathrm{qe}},
\end{equation}
where the quantum efficiency $\eta_\mathrm{qe}$~(=0.84) is linked to the photodiode responsivity $R$ (=1~A/W) via
\begin{equation}
R = \frac{e\eta_\mathrm{qe}}{\hbar \omega_o}.
\end{equation}
In our balanced detection scheme, we consider the shot noise of the difference photocurrent of the two detectors. Since photon annihilation at the two detectors is uncorrelated, the total shot noise is given by the incoherent sum of the two individual power-spectral-densities, such that
\begin{equation}\label{EqSPPSN}
S_{PP}^{SN} = \frac{2\hbar\omega_o P_{\mathrm{tot}}}{\eta_\mathrm{qe}}
\end{equation}
with $P_\mathrm{tot} = P_\mathrm{det1} + P_\mathrm{det2}$ being the sum of the individual powers hitting the two photodiodes. In our balanced detection scheme, $P_\mathrm{det1} = P_\mathrm{det2}$ and $P_\mathrm{tot} = 2 T \eta_\mathrm{in} P_\mathrm{in}$. While the balanced detection scheme used in our experiment is beneficial towards the suppression of technical laser amplitude noise, it hence comes with the disadvantage of introducing more shot noise into the system. In this experiment, the noise-equivalent power corresponding to shot noise is $6.1\ \mathrm{pW/\sqrt{Hz}}$. The noise-equivalent acceleration corresponding to this noise background is given by
\begin{eqnarray}
a_\mathrm{SN}(\omega) = \sqrt{S_{aa}^{SN}} & = & \frac{1}{\left| \frac{\mathrm{d}T}{\mathrm{d}\Delta} \right| \eta_\mathrm{in} P_\mathrm{in} g_\mathrm{OM} } \frac{1}{\left| \chi(\omega) \right|} \sqrt{S_{PP}^{SN}} \\
 & = & \frac{\omega_0}{(1-T_d) Q_o g_\mathrm{OM}} \, \frac{1}{|\chi{(\omega)}|} \sqrt{\frac{2\hbar\omega_o(1+T_d)}{\eta_\mathrm{qe} \eta_\mathrm{in} P_\mathrm{in}}}. \label{EqNoiseTransd}
\end{eqnarray}
With the values given above, this yields $a_\mathrm{SN} = 8.9\ \mathrm{\mu g/\sqrt{Hz}}$ around DC.

\subsection{Detector noise}
The electronic detector noise is usually quantified by the noise-equivalent-power (NEP), which for the Newport 2117 detector and the transimpedance gain setting we use is on the order of $2.8\ \mathrm{pW/\sqrt{Hz}}$. The optical noise power-spectral-density then is
\begin{equation}
S_{PP}^{NEP}(\omega) = \mathrm{NEP}^2.
\end{equation}
In analogy to eq.~(\ref{EqNoiseTransd}), the NEA corresponding to electronic detector noise can be derived as
\begin{equation}\label{EqaNEP}
a_\mathrm{NEP} = \frac{\omega_o}{(1-T_d)Q_og_\mathrm{OM} \eta_\mathrm{in} P_\mathrm{in}} \frac{1}{|\chi(\omega)|} \mathrm{NEP}.
\end{equation}
Here, this is found to be $a_\mathrm{NEP} = 4.1\ \mathrm{\mu g/\sqrt{Hz}}$ at DC.

\subsection{Backaction noise}\label{SecBackactionNoise}
The extra noise $a_\mathrm{add}$ added by the optical field mentioned above arises from optical noise that exerts a random force on the mechanical oscillator via radiation pressure. The optical noise arises from classical amplitude noise and from intrinsic shot noise. In the following, we consider only quantum back-action noise $a_\mathrm{BA}$ arising optical shot noise. With $\hbar g_\mathrm{OM}$ being the force exerted per photon and for $n_\mathrm{cav}$ photons in the cavity, the random acceleration created by optomechanical back-action has a power spectral density of~\cite{clerk_introduction_2010}
\begin{equation}
S_{aa}^{BA} = 2\frac{(\hbar g_\mathrm{OM})^2}{m^2} n_\mathrm{cav} \frac{4}{\kappa},
\end{equation}
resulting in a noise-equivalent acceleration of $a_\mathrm{BA} = \sqrt{S_{aa}^{BA}} = 5.6\ \mathrm{ng/\sqrt{Hz}}$.  Here, owing to the low quality factor of the optical cavity and the low mechanical frequency, the shot noise radiation pressure force is approximately white noise for frequencies of relevance near the mechanical frequency.  Note also that we are using single-sided PSDs, hence double the value of the (approximately) symmetric double-sided PSD. This value is much smaller than the acceleration noise created by the other sources discussed previously. The frequency-dependent displacement noise created by quantum back-action is
\begin{equation}
S_{xx}^{BA}(\omega) = 2\left(\frac{2\hbar g_\mathrm{OM}}{m}\right)^2 \frac{n_\mathrm{cav}}{\kappa} |\chi(\omega)|^2
\end{equation}
On the mechanical resonance, and using eq.~(\ref{Eqncav2}), this yields
\begin{equation}
S_{xx}^{BA}(\omega_m) = 2\left( \frac{ 2\hbar g_\mathrm{OM} Q_m }{ m \omega_m^2 } \right)^2 \frac{ 1-\sqrt{T_d} }{ \kappa^2 } \frac{ P_\mathrm{in} }{ \hbar \omega_c },
\end{equation}
resulting in $\sqrt{S_{xx}^{BA}(\omega_m)} = 23\ \mathrm{fm/\sqrt{Hz}}$ for the device and experimental conditions described in Figs.~\ref{Fig2} and \ref{Fig3} ($P_\mathrm{in}=116$~$\mu$W). This should be compared to the fundamental standard quantum limited displacement noise given by
\begin{eqnarray}
S_{xx}^{SQL}(\omega) & = & S_{aa}^{SQL} |\chi(\omega)|^2 \\
 & = & \frac{ 2\hbar \omega_m \gamma_m }{ m } |\chi(\omega)|^2,
\end{eqnarray}
which on resonance has the simple form
\begin{equation}
S_{xx}^{SQL}(\omega_m) = x_\mathrm{zpm}^2 \frac{ 4 }{ \gamma_m },
\end{equation}
with the zero-point motion given by
\begin{equation}
x_\mathrm{zpm} = \sqrt{\frac{\hbar}{2 m \omega_m}}.
\end{equation}
For the device and experimental conditions described in Figs.~\ref{Fig2} and \ref{Fig3}, this yields an on-resonance SQL of $\sqrt{S_{xx}^{SQL}} = 2.8\ \mathrm{fm/\sqrt{Hz}}$.

\subsection{General discussion}

\begin{figure}[t]
\includegraphics{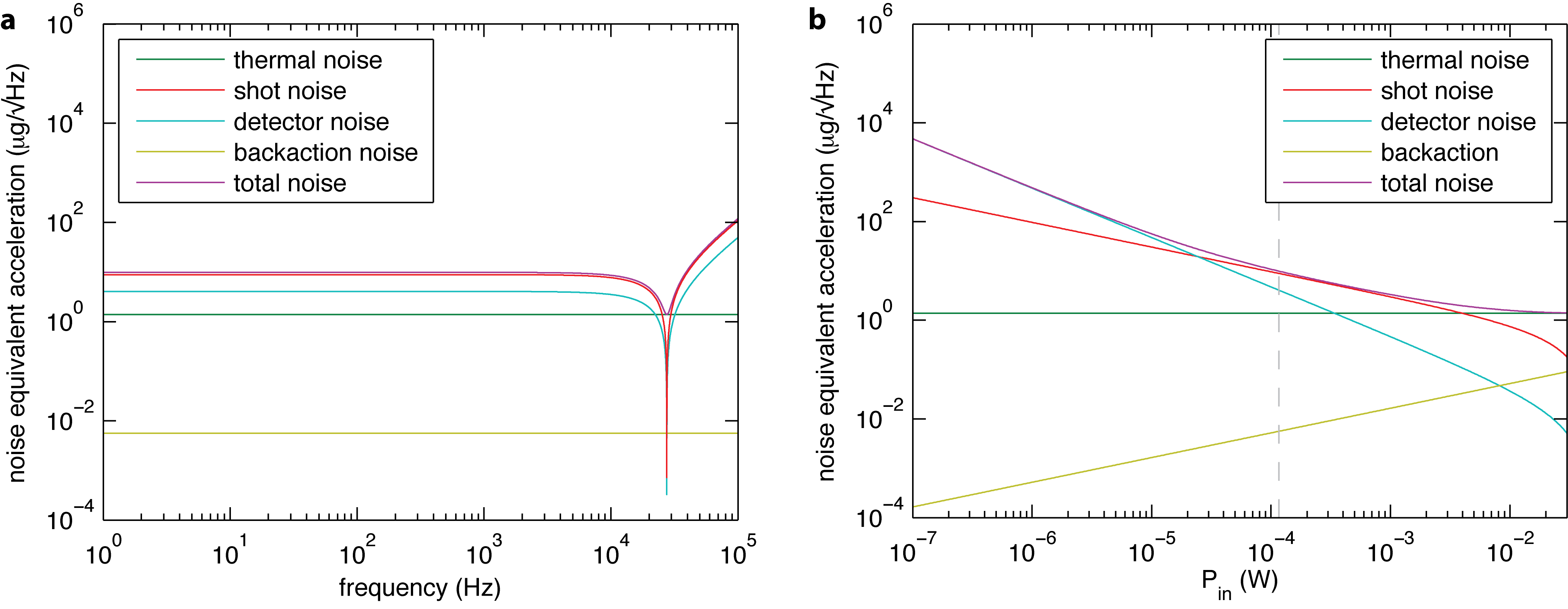}
\caption{\label{AccPaper_SI_FigS1} \textbf{Frequency response of acceleration noise and power-scaling.}
	\textbf{a}, Frequency dependent contributions of different noise sources to the noise-equivalent-acceleration (NEA). For this calculation, we used the same parameters as for the device shown in Fig.~\ref{Fig2}.
	\textbf{b}, Dependence of the DC-NEAs as function of incident laser power. The dashed vertical line indicates the optical power used in measurements of Figs.~\ref{Fig2}b and \ref{Fig3}.}
\end{figure}
The dashed lines in Fig.~\ref{Fig2}b show the contributions of these noise terms to the PSD of the balanced photo-detector output, where we neglected back-action noise. Figure~\ref{AccPaper_SI_FigS1}a shows the corresponding frequency-dependent noise-equivalent acceleration values corresponding to the different noise terms for the device studied here. Here, we include $a_\mathrm{BA}$, which can be seen to only contribute negligibly to the NEA of the device. While $a_\mathrm{th}$ (green) and $a_\mathrm{BA}$ (gold) are frequency-independent, the NEAs of photon shot noise $a_\mathrm{SN}$ (red) and electronic detector noise $a_\mathrm{NEP}$ (cyan) are colored by the frequency-dependent response of the oscillator $\chi(\omega)$.

While thermal noise arises as a fundamental property of a mechanical oscillator in contact with a heat bath at temperature $T$, the contributions of shot noise and detector noise are dependent on the efficiency of the optomechanical transduction mechanism. From eq.~(\ref{EqNoiseTransd}) and eq.~(\ref{EqSPPSN}), one can see that $a_\mathrm{SN} \propto P_\mathrm{in}^{-1/2}$, while from eq.~(\ref{EqaNEP}) it follows that $a_\mathrm{NEP} \propto P_\mathrm{in}^{-1}$. Similarly, back action noise scales with the square-root of the number of photons in the cavity: $a_\mathrm{BA} \propto P_\mathrm{in}^{1/2}$. For illustration, Fig.~\ref{AccPaper_SI_FigS1}b shows the relative contributions of the individual noise terms at DC ($\omega=0$) as function of incident power. Here, we include the effects of thermo-optomechanical back-action on the mechanical susceptibility, as discussed in appendix~\ref{SecBackactTheory}. The thermal noise background is not affected by cooling of the mechanical mode, since it follows from eq.~(\ref{Eqath_SI}) that $a^2_\mathrm{th} \propto T_\mathrm{eff} \gamma_m = \mathrm{const}$ under back-action damping/cooling of the mechanical mode. The roll-off of shot noise and detector noise for pump powers above 10~mW arises from the decrease of the mechanical mode frequency $\omega'_m$ due to the optomechanical spring effect. This results in an increase of the DC acceleration sensitivity $\chi(0) = 1/{\omega'_m}^2$ and thereby a reduction of the corresponding acceleration noise floors according to eqs.~(\ref{EqNoiseTransd}) and (\ref{EqaNEP}).

As mentioned previously, for the power used in the experiment, the NEA is limited by photon shot noise. For two orders of magnitude higher pump powers, the NEA starts being dominated by thermal noise of the test-mass oscillator. Alternatively, according to eq.~(\ref{EqNoiseTransd}), thermal-noise limited detection can be achieved by increasing $g_\mathrm{OM}$ by one order of magnitude.

\section{Linear Dynamic Range}\label{SecLinDynRange}

\begin{figure}[h]
\includegraphics{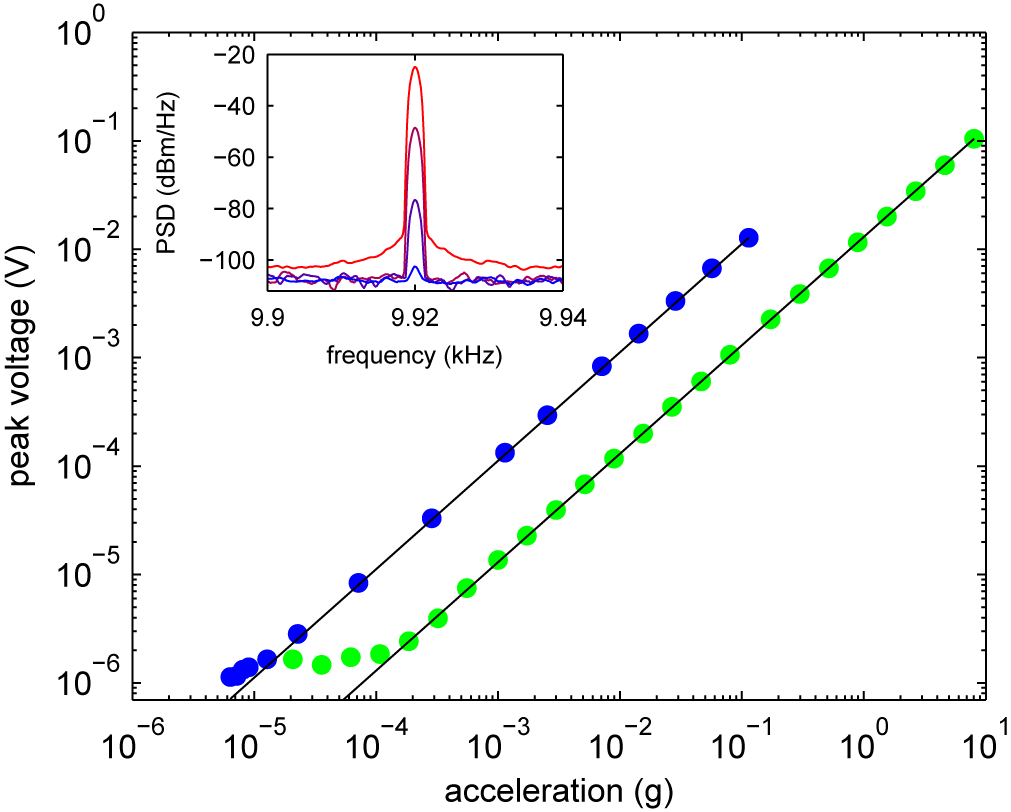}
\caption{\label{AccPaper_SI_FigS4}
	\textbf{Demonstration of large linear dynamic range of typical devices.}
	While varying the amplitude of the acceleration applied with the calibrated shake table at 9.92~kHz, we measure the optical signal transduced via the mechanical mode.  The blue bullets show the transduced signal of the device presented in the text using the voltage corresponding to the peak height of the modulation tone on the ESA spectrum, which exhibits linear response over 40~dB.  The inset shows the corresponding PSD spectra from the ESA for modulation tones between 0.1~g and $12.8\ \mathrm{\mu g}$, taken at a resolution bandwidth of 1~Hz.  The green bullets show data obtained from a different device with a larger thermal noise background but very similar optomechanical coupling using the lock-in scheme depicted in Fig.~\ref{Fig2}a, which exhibits linear response over 49~dB -- limited by the maximum voltage output of our function generator, which corresponds to an acceleration of 8~g.  The black lines are linear fits to the data.}
\end{figure}

A key requirement for any inertial sensor is linear response over a reasonable dynamic range.  To check the linearity of the response of the accelerometer presented in the text, we varied the amplitude of a sinusoidal signal sent to the shear piezo at 9.92 ~kHz and recorded the voltage corresponding to the peak height of the transduced modulation tone -- shown in blue bullets in Fig.~\ref{AccPaper_SI_FigS4}.  The sensor behaves linearly over a dynamic range of 41 ~dB, with the tone vanishing into the shot noise floor for an applied acceleration of $\approx 10\ \mathrm{\mu g}$ at a resolution bandwidth of 1~Hz.  The green bullets in Fig.~\ref{AccPaper_SI_FigS4} show data from a different device with similar geometry but slightly lower mechanical $Q$-factor, which exhibits a linear response over 49~dB.  This particular measurement was limited by the maximum output voltage of the function generator.  Ultimately, however, the linear dynamic range ends when motion of the test mass shifts the optical resonance by a magnitude comparable to the optical cavity linewidth.  For this device, this is expected to occur for accelerations of $\sim 50\ \mathrm{g}$ for frequencies below $\sim 25\ \mathrm{kHz}$.

\end{document}